\documentclass[aps,twocolumn,groupedaddress]{revtex4-1}

\usepackage{graphicx}
\usepackage{dcolumn}
\usepackage{bm}
\usepackage{amsfonts}
\usepackage{amsmath}
\usepackage{amssymb}
\usepackage{xcolor}
\usepackage{enumerate}
\usepackage{soul}

\usepackage{hyperref}
\hypersetup{
    colorlinks,%
    citecolor=blue,%
    linkcolor=blue,%
    urlcolor=blue
}

\setcitestyle{super}
\newcommand{\can}{Ca$_2$N}

\begin{document}

\title{Spin-polarized nearly-free electron channels on the \can\, electrenes }

\author{Pedro H. Souza$^1$}
\email{psouza8628@gmail.com}

\author{Jos\'e E. Padilha$^2$}
\email{jose.padilha@ufpr.br}

\author{Roberto H. Miwa$^1$}
\email{hiroki@ufu.br}

\affiliation{$^1$Instituto de F\'isica, Universidade Federal de Uberl\^andia,
        C.P. 593, 38400-902, Uberl\^andia, MG, Brazil}
        
\affiliation{$^2$Campus Avan\c cado Jandaia do Sul, Universidade Federal do 
Paran\'a, \\ 86900-000, Jandaia do Sul, PR, Brazil.}

  
\begin{abstract}
\vspace{3mm}
\begin{center}
 {\bf ABSTRACT}
\end{center}

Two-dimensional (2D) materials combined with the presence of surface nearly-free electrons (NFE) have been considered quite interesting platforms to be exploited for the development of 2D electronic devices. Further incorporation of foreign elements adds a new degree of freedom to engineer the electronic as well as the magnetic properties of 2D materials. Here we have performed an {\it ab-initio} study of \can\, electrenes fully (i.e., both sides) adsorbed by hydrogen (H/\can/H) and fluorine (F/\can/F) atoms. The NFE states are suppressed in these systems, followed by the appearance of a net magnetic moment localized in the nitrogen atoms intercalated by the hydrogenated or fluorinated calcium layers. In the sequence, we have proposed lateral heterostructures combining the H/\can/H or F/\can/F regions with pristine \can\, electrenes [(\can)(X/\can/X), with X\,=\,H or F]. We found that the magnetic moment of the hydrogenated or fluorinated regions promotes the emergence of spin-polarized NFE states confined along the pristine  (\can) stripes. Further electronic transport calculations reveal that the (X/\can/X) regions act as spin-dependent scattering centers,  spin-filters. We believe that these findings make an important contribution to the development of spintronic devices based on 2D electrides.

\end{abstract}
 
\maketitle

\section{Introduction}

Electrides are a class of ionic crystals in which there are excess valence electrons that behave as anions and don't belong in an atomic orbital \cite{oh2016evidence, sui2019prediction, lee2013dicalcium}. There are some electrides that crystallize in a layered structure and these anionic electrons are located in the interlayer gap forming a 2D electron gas. As a consequence, these materials offer properties such as low work functions, high electrical conductivity, and rich surface chemistries \cite{druffel2017electrons}. The layered structure of electrides permits the synthesis of 2D electrides by an exfoliation process, once that interlayers bonding is van der Waals (vdW) \cite{wang2018ultralow}. The electride monolayer, often referred as \textit{electrene}, has been experimentally obtained with electride samples of Ca$_{2}$N and Y$_{2}$C. In these samples, nearly free electrons (NFE) are situated on both sides of the monolayer
\cite{druffel2016experimental,dale2017ionic}.

Some studies show that electrides are interesting materials that can have different physical properties and applications such as topological phases\cite{hirayama2018electrides, huang2018topological, park2018first, zhu2019computational, zhang2018intermetallic, zhang2019topological, liu2019theoretical}, great potential as an anode to batteries \cite{hou2016two, chen2017multilayered, 2018determination}, metal-semiconductor transition under pressure\cite{tang2018metal} and doping effects on vdW heterostructures with graphene \cite{inoshita2017probing, choi2021electronic}.   

The presence of anionic electrons located on both sides of electrenes makes the surface be reactive. Furthermore, the structure resembling MXenes facilitates the functionalization process, potentially leading to varied electronic and magnetic properties compared to the pristine structure\cite{wu2020intriguing}. Recently, we have shown that the full oxidation on one side of monolayer \can\/ and both sides of bilayer \can\/ induce a structural transition from hexagonal to tetragonal, followed by the emergence of half-metallic channels, near the Fermi level, shielded from the environment conditions by the CaO (cover) layers\cite{souza2020structural}. Moreover, the \can\ functionalization via Hydrogen (H) induces a metalic-semiconductor transition when the H is adsorbed on one side of the \can\ surface, and when H is adsorbed on both sides of the surface the functionalized \can\/ becomes a half-metal \cite{hidrox}. In addition, the process of functionalization enables some applications of the electrenes, such as in batteries through adsorption of hydroxyl \cite{wang2019first}. 

The design of heterostructures, guided by a strategic combination of materials with suitable properties, has been the subject of intense research addressing the development of electronic devices that meet the current demand. Such a combination can be done, in most cases, by vertical  or lateral (in-plane) stacking  of different materials like WS$_2$/MoS$_2$\cite{gong2014vertical,fang2018chemical} or graphene/boron-nitride\cite{ci2010atomic}. In particular, lateral heterostructures based on 2D platforms can also be realized by combining distinct structural phases of the same material\cite{liu2018intermixing,silvestre2019electronic} or through the selective incorporation (adsorption) of foreign elements, giving rise nanoroads with magnetic properties in graphene oxide, and  diamane oxide films\cite{khabibrakhmanov2022electronic,varlamova2022diamane}.

In this work, we performed an {\it ab-initio} study of the energetic stability and electronic properties of fully hydrogenated (H/\can/H) and fluorinated (F/\can/F) electrenes. We found that both systems are characterized by (i) the suppression of the NFE states on the \can\, surface, and (ii) the a net magnetic moment localized in the nitrogen atoms. In the sequence, we focused on the lateral heterostructures composed by a combination of H/\can/H or F/\can/F regions (or stripes) separated by pristine \can\, stripes [(\can)(X/\can/X), with X\,=\,H or F]. Our findings reveal the emergence of channels of spin-polarized NFE surface states on the pristine \can\, regions, which are induced by the X/\can/X regions. Further electronic transport calculations show that the electron scattering rate is spin-dependent in the (X/\can/X) regions, thus acting as a spin-filters. 

\section{Computational details}
The calculations were performed by using the density functional theory (DFT)\cite{hk}, as implemented in the Vienna Ab initio Simulation Package (VASP)\cite{vasp1,vasp2}. The exchange-correlation functional was considered the gradient generalized approximation of Perdew-Burke-Ernzerhof (GGA-PBE)\cite{pbe}.   
The atomic positions of the primitive hexagonal cell of \can\/, H/\can/H, and F/\can/F were relaxed until the residual forces were converged to within 5\,meV/\AA\, and the structural relaxation (variable-cell) was performed within a pressure convergence of 0.05 Kbar  with a Brillouin zone sampling of the 8$\times$8$\times$1 k-point mesh\cite{kpoints}. In the lateral heterostructures we used supercells  with a vacuum region of 10\,\AA\, and considered an energy cutoff of 500\,eV for the plane wave basis set. The Brillouin zone was sampled using a 6$\times$9$\times$1 and 9$\times$6$\times$1 k-point mesh\cite{kpoints}, for zigzag and armchair directions, respectively, and more structural supercell details are described throughout the text.  

The electronic transmission probability (\textit{T}) was calculated within the non-equilibrium Green's Functions (NEGF) formalism using the DFT Hamiltonian and implemented in the Siesta and TranSiesta codes\cite{siesta,transiesta}. The Kohn-Sham orbitals were expanded in a linear combination of numerical pseudo-atomic orbitals, using a single-zeta basis set including polarization functions\cite{artacho1999sanchez}. The BZ samplings were performed using two different sets of k-point meshes, 1$\times$20$\times$300 according to the electronic transport directions. The total  transmission probability of electrons with energy $E$ and bias voltage $V$, $T(E, V)$, from the left electrode to reach the right electrode passing through the scattering region is given by, 
$$T\left(E \right) = Tr \left[\Gamma_{\mathrm R}\left(E,V\right) G^{\mathrm R}\left(E,V\right) \Gamma_{\mathrm L}\left(E,V\right) G^{\mathrm A}\left(E,V\right) \right], $$
where $\Gamma_{L,(R)} \left(E,V\right)$ is the coupling with the left and right electrodes and $G^{R,(A)}$ is the retarded (advanced) Green function matrix of the scattering region. 

\section{Results and Discussions}

In Fig.\,\ref{pristines}(a1) we show the structural model and the electronic band structure of pristine \can\, monolayer (ML). We have considered a tetragonal unit-cell with (equilibrium) lattice vectors of  $|{\bf a}|$\,=\,3.55\,\AA, and $|{\bf b}|$\,=\,6.17\,\AA\,. The electronic band structure, Figs.\,\ref{pristines}(a2) and (a3), is characterized by (i) the formation of  parabolic metallic bands at the $\Gamma$ point, a signature of the emergence of NFE on the ML surface, and (ii) a set of fully occupied energy bands, between  2 and 4\,eV below the Fermi level ($E_\text{F}$), attributed to the N-$2p$ orbitals. 


\subsection{Hydrogenation and Fluorination}

\begin{figure}
    \includegraphics[width=8cm]{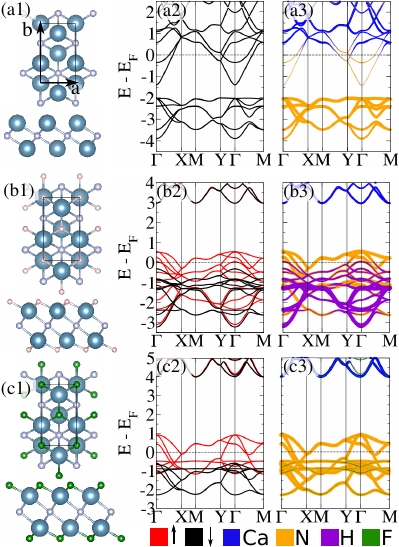}
    \caption{Structural model of \can\ pristine (a1), fully hydrogenated  (b1), and fully fluorinated (c1); their respective electronic band structure  (a2,b2,c2) and the projection of the energy bands onto atomic orbitals (a3,b3,c3). The balls represent Calcium atoms (blue), Nitrogen (grey), Hydrogen (pink), and Fluorine (green). }
    \label{pristines}
\end{figure}

A promising method to regulate or fine-tune the mechanical and electrical features of 2D systems is hydrogenation or fluorination. In the case of graphene as the 2D platform, semiconductor systems are produced by the complete adsorption of H or F atoms, resulting in graphane or fluorgraphene\cite{leenaerts2010first}.

\begin{figure*}
 \centering
    \includegraphics[width=15.5cm]{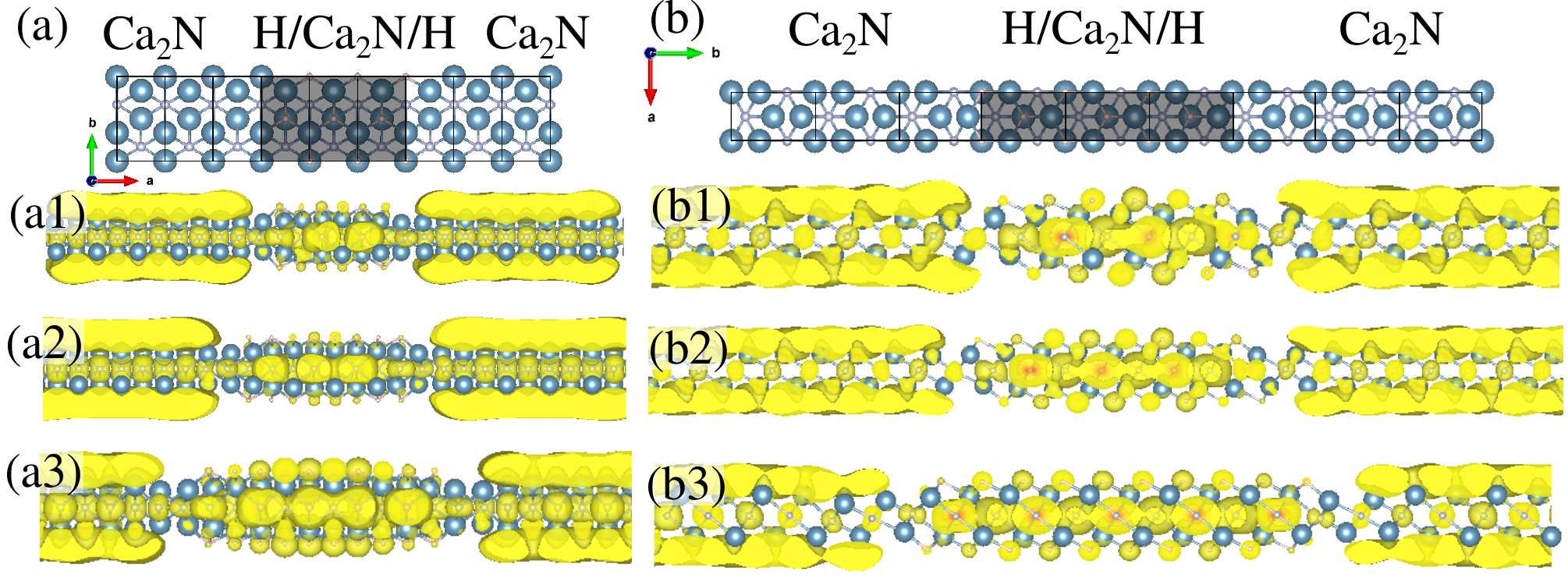}
    \caption{Structural model of the zigzag (a) and armchair (b) (\can)$_6$(H/\can/H)$_{n}$ lateral heterostructures, and the partial charge density distribution near the Fermi level, $E_F\pm 0.2$\,eV for $n$\,=\,3 (a1)-(b1),  $n$\,=\,4 (a2)-(b2), and  $n$\,=\,5 (a3)-(b3). }
   \label{fig:hetero}
\end{figure*}

In Figs.\,\ref{pristines}(b1) and (c1) we present the lowest energy geometries of the fully hydrogenated and fluorinated \can\,ML (hereinafter labeled as H/\can/H and F/\can/F), which are characterized by the (H/F) adatoms aligned with the Ca atoms at the opposite side of the ML\cite{hidrox}.  Our results of the equilibrium lattice constants ($|{\bf a}|$ and $|{\bf b}|$), see Table\,\ref{table:energy}, reveal that the \can\,ML lattice is slightly shrinked by about 2\,\%. The formation energy ($E^f$) and binding energy ($E^b$) of X/\can/X (with X\,=\,H or F) is defined as the total energy difference between the final system and the sum of the total energies of the separated components, \can\,ML and X$_2$ molecule, and  X atom, respectively. Negative values indicate exothermic processes. Our result of $E^b$ for H/\can/H agrees with that obtained by Qiu {\it et al.}\cite{hidrox}. On the other hand, it is worth noting that in $E^f$ we are comparing the final system, X/\can/X, with respect to the energetically stable molecular systems, X$_2$. Thus, our results of $E^f$ allow us to infer that F/\can/F is substantially more stable than  H/\can/H. Indeed, a similar stability picture has been predicted in graphene, i.e. graphane versus fluorgraphene\cite{leenaerts2010first}.

\begin{figure}
 \centering
    \includegraphics[width=7cm]{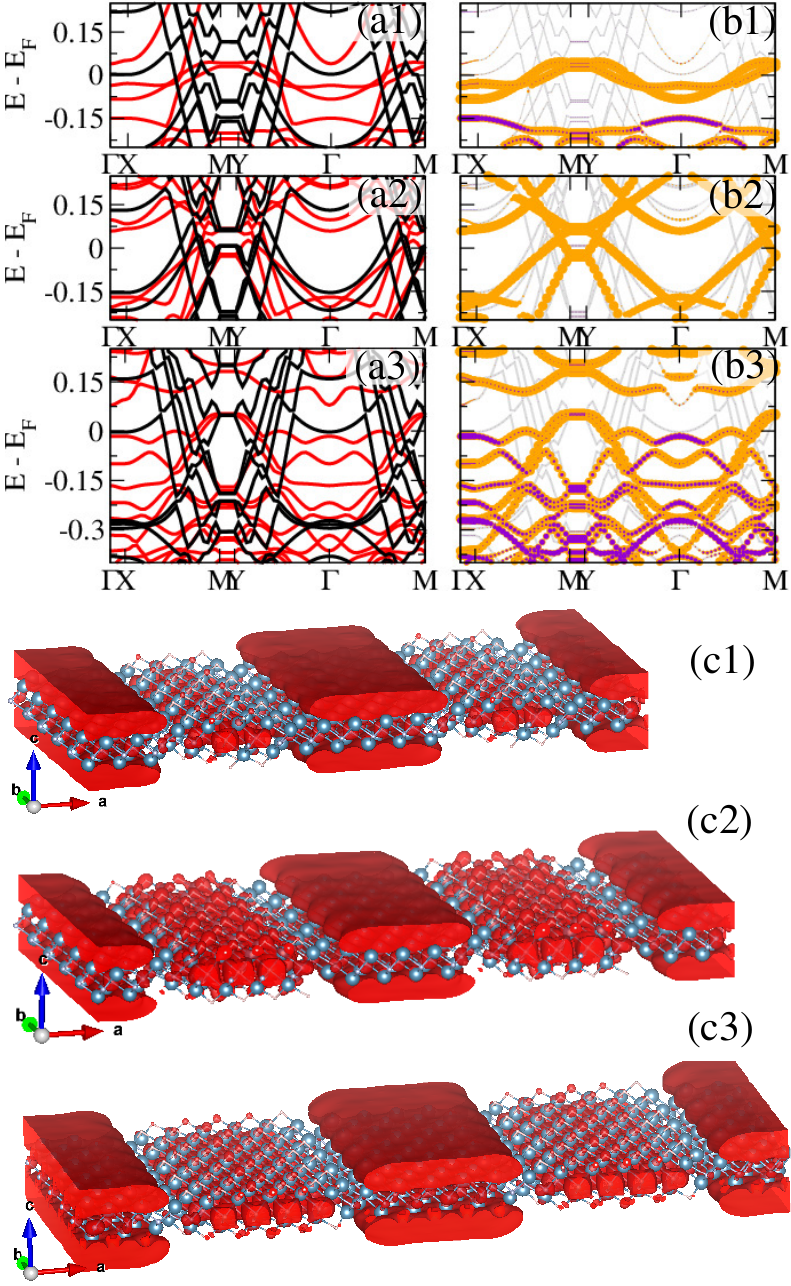}
    \caption{(a) Spin-polarized electronic band structure, and (b) the respective projection of hidrogenated region of (\can)$_6$(H/\can/H)$_n$, for $n$\,=\,3 (a1)-(b1), 4 (a2)-(b2), and 5 (a3)-(b3). (c) Spin-density ($\Delta\rho^{\uparrow\downarrow}$) of  (\can)$_6$(H/\can/H)$_n$ for $n$\,=\,3 (c1), 4 (c2) and 5 (c3).}
   \label{fig:cell}
\end{figure}

\begin{table}[!htb]
\caption{Binding energy ($E^b$) and formation energy ($E^f$) in eV/adatom, and the equilibrium lattice constants,  $|{\bf a}|$ and $|{\bf b}|$  in \AA.}
\begin{ruledtabular}
\begin{tabular}{ccccc}
          & $E^b$  &  $E^f$      & $|{\bf a}|$ & $|{\bf b}|$  \\
            \hline
\can-H     &  $-2.62$  & $-0.43$ &  3.50     & 6.06      \\
\can-F     & $-4.19$  & $-3.71$       & 3.48      & 6.03 \\
\end{tabular}
\end{ruledtabular}
\label{table:energy} 
\end{table}

Upon the formation of Ca-H or Ca-F chemical bonds,  we find that (i) the anionic electrons on the \can\, surface has been suppressed, as well as the NFE parabolic bands crossing the Fermi level, as depicted in Figs.\,\ref{pristines}(b2)-(b3) and (c2)-(c3);  followed by (ii)  an increase of the work function ($\Phi$) from 3.6 eV (pristine \can\,ML) to 5.44 and 5.91\,eV in H/\can/H and F/\can/F respectively; and (iii) the emergence of a net magnetic moment of 0.63\,$\mu_\text{B}$/N. In both systems, the magnetization is dictated by the unpaired states near the Fermi level, resulting in half-metallic systems mediated by the planar N-$2p_{xy}$ orbitals.  At the ground state configuration, X/\can/X presents a ferromagnetic (FM) phase,  which is in agreement with Qiu {\it et al.},  and characterized by an energy gain, $\Delta E^\text{mag}=E^\text{FM}-E^\text{NM}$, of $-200$ and $-170$\,meV/unit-cell, for X\,=\,H and F, respectively. In Ref.\,\cite{hidrox}, the  authors obtained  $\Delta E^\text{mag}$ of $-157$\,meV/unit-cell in H/\can/H.

\subsection{Lateral Heterostructures}







A promising route for creating electronic (nano)devices is the construction of 1D structures at the atomic level on 2D platforms.  For instance, the creation of nanodots and linear stripes on graphene as a result of the recent manipulation of hydrogen adatoms has produced 0D and 1D electronic confinement phenomena.\cite{cortes2020quantum} It is worth noting that, more than ten years ago, several works have already predicted such electronic confinement and the emergence of magnetism in 1D nanostructures on graphane\cite{chernozatonskii2007two,wu2009materials,singh2009electronics}, and more recently on graphene oxide\cite{khabibrakhmanov2022electronic}. 

As discussed above,  the pristine \can\,ML is non-magnetic and has NFE states localized on the surface. In contrast, H/\can/H and F/\can/F are characterized by the emergence of magnetic moment localized on the nitrogen atoms, resulting in half-metallic states. In view of these results, in the sequence, we investigated the electronic and  magnetic properties of the  (\can)(X/\can/X) lateral heterostructures, characterized by pristine \can\, ML intercalated by periodic stripes of hydrogenated or fluorinated regions, (\can)(X/\can/X)$_n$, with  $n$=3, 4, and 5.

We have studied (\can)(X/\can/X)$_n$ heterostructures composed by six unit cells of \can, namely (\can)$_6$(X/\can/X)$_n$, forming armchair (ARM) [Fig.\,\ref{fig:hetero}(a)] and zigzag (ZZ) [Fig.\,\ref{fig:hetero}(b)] interfaces between the (\can)$_6$ and (X/\can/X)$_3$ stripes. We found that the electronic states close to the Fermi level, $E_\text{F}\pm 0.2$\,eV, are predominantly localized on the (i) nitrogen atoms intercalated by hydrogenated/fluorinated calcium layers and (ii) on the surface of the non-hydrogenated \can\, regions, as depicted in Figs.\,\ref{fig:hetero}(a1)-(a3) and (b1)-(b3) for X\,=\,H. These findings suggest the emergence of NFE surface states propagating along the (\can)$_6$ stripes, confined by the hydrogenated (H/\can/H)$_n$ walls. Similar results were obtained in the fluorinated heterostructures, (\can)$_6$(F/\can/F)$_n$. 

Total energy comparisons between the ARM and ZZ interfaces, $\Delta E=E_\text{ARM}-E_\text{ZZ}$, reveal an energetic preference for the ARM interface, Table\,\ref{table:preference}. As expected, the role played by the orientation of the edge interface becomes less important as the hydrogenated or fluorinated region increases, i.e. $\Delta E\rightarrow 0$ for larger values of $n$. 

\begin{table}[!htb]
\caption{Total energy difference ($\Delta E$) between the ZZ and ARM (\can)$_6$(X/\can/X)$_n$ heterostructures, X\,=\,H and F.}
\begin{ruledtabular}
\begin{tabular}{rc}
 (\can)$_6$(X/\can/X)$_n$ & $\Delta E=E_\text{ARM}-E_\text{ZZ}$    \\ 
\hline
(\can)$_6$(H/\can/H)$_3$  &  -0.144 eV/H      \\
(\can)$_6$(H/\can/H)$_4$&   -0.101 eV/H  \\  
(\can)$_6$(H/\can/H)$_5$   &  -0.073 eV/H \\ 
\hline 
(\can)$_6$(F/\can/F)$_3$  &  -0.136 eV/F   \\
(\can)$_6$(F/\can/F)$_4$  &  -0.103 eV/F   \\  
(\can)$_6$(F/\can/F)$_5$   &  -0.069 eV/F     
\end{tabular}   
\end{ruledtabular}
\label{table:preference} 
\end{table}

\begin{figure}
    \includegraphics[width=8cm]{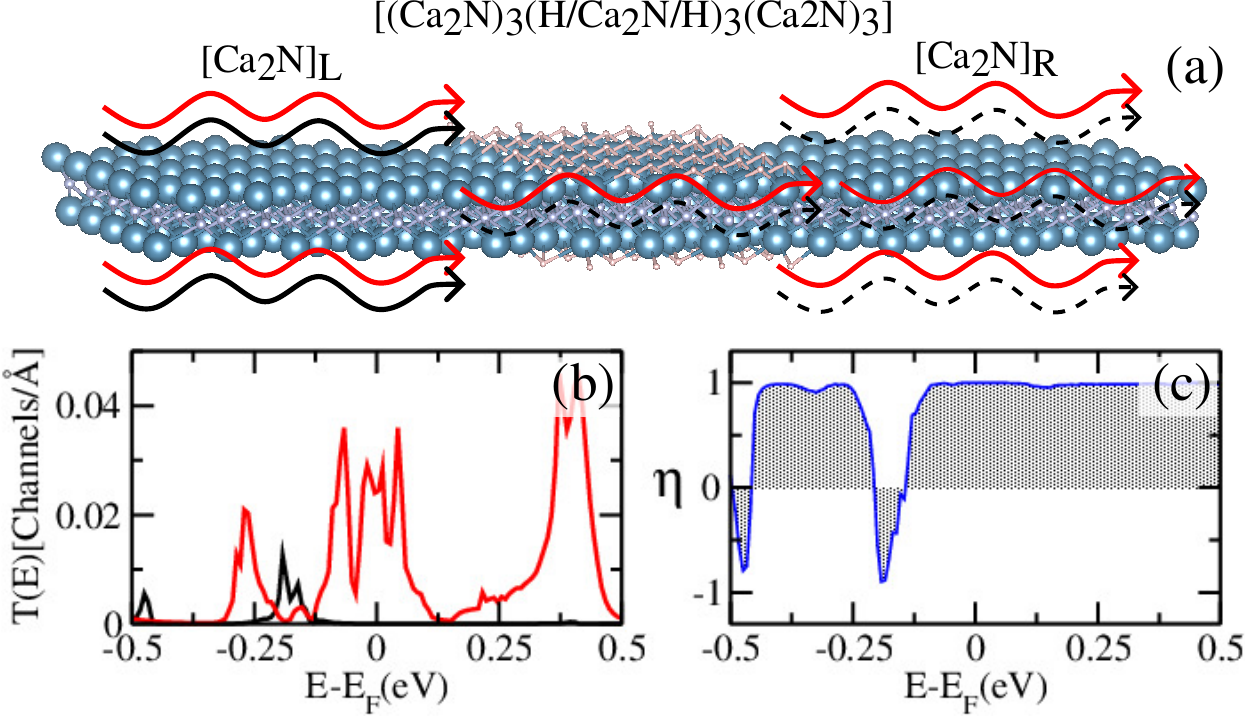}
    \caption{(a) Schematic model of the setup used for the electronic transport calculation, (b) transmittance  [$T(E)$] of the spin-up [$T^\uparrow(E)$] (red-lines) and spin-down [$T^\downarrow(E)$] (black-lines) channels, and (c) spin-anisotropy of the transmittance  [$\eta(E)$]. \label{setup}} 
\end{figure}

In the sequence, we will focus on the electronic properties of the armchair (\can)$_6$/(H/\can/H)$_n$ heterostructures.  The presence of a net magnetic moment of the nitrogen atoms sandwiched by the hydrogenated Ca layers, H/\can/H, leads to the formation of spin-polarized bands, indicated by the red (spin-up) and black (spin-down) solid lines in  Figs.\,\ref{fig:cell}(a1)-(a3), for $n$\,=\,3, 4, and 5. Indeed, further orbital projected electronic band structure on the hydrogenated region [Fig.\,\ref{fig:cell}(b1)-(b3)] reveals that the spin-up N-$2p$ orbitals bring an important contribution to the spin-polarized states near the Fermi level. It is noticeable the predominance of spin-up metallic bands for wave-vectors parallel (Y-$\Gamma$ and X-M directions) and quasi-parallel ($\Gamma$-M direction) to the (\can)$_6$/(H/\can/H)$_n$ stripes. Meanwhile, along the $\Gamma$-X and Y-M directions, the electronic states are characterized by dispersionless energy bands resulting from the electronic confinement perpendicularly to the stripes. Further spin-density distributions, $\Delta\rho^{\uparrow\downarrow}=\rho^\uparrow-\rho^\downarrow$, depicted in Figs.\,\ref{fig:cell}(c1)-(c3), confirm the formation of 1D spin-polarized NFE states spreading out on the  (\can)$_6$ surface confined by  (H/\can/H)$_n$. We found similar results in the fluorinated systems, (\can)$_6$(F/\can/F)$_n$, Fig.\,\ref{apen2} in the appendix. Those findings suggest the emergence of spin-polarized electronic transport channels based on the NFE along the \can\, confined by hydrogenated or fluorinated stripes.

On the other hand, it is expected that the NFE states are no longer spin-polarized far from the hydrogenated or fluorinated stripes,   so it is worth examining the role played by the hydrogenated stripes in the scattering process of the non-spin-polarized NFE states. In this case, as a test of concept, we have considered non-spin-polarized (pristine) \can\,  leads connected to the (\can)$_3$(H/\can/H)$_3$(\can)$_3$ lateral heterostructure, \\

[\can]$_\text{L}$--(\can)$_3$/(H/\can/H)$_3$/\can)$_3$--[\can]$_\text{R}$.\\ 

[\can]$_\text{L}$ and [\can]$_\text{R}$ indicate the left and right leads, respectively, as schematically shown in Fig.\,\ref{setup}(a). Our results of the transmittance, $T(E)$, close to the Fermi level [Fig.\,\ref{setup}(b)] demonstrates the dominance of the spin-up transmission channels [$T^\uparrow(E)$], leading to a net spin-polarized current [Fig.\,\ref{setup}(b)]. Such a spin-anisotropy [$\eta(E)$] in the electronic transport properties can be measured as,
$$
\eta(E)=\frac{T^\uparrow(E) - T^\downarrow(E)}{T^\uparrow(E) + T^\downarrow(E)}.
$$
Our results of $\eta(E)$, presented in Fig.\,\ref{setup}(c), show the emergence of a nearly full spin-up current [$\eta(E)\approx 1$] at the limit of low bias voltage, $|E-E_\text{F}|\leq 0.12$\,eV; followed by a contribution of $T^\downarrow$ upon further increase of the bias voltage, due to the presence of spin-down channels for $E-E_\text{F}\approx -0.15$\,eV. Such dominance of spin-up transmittance can be attributed to (i) the spin polarization of the (\can)(H/\can/H)(\can) contact region, and (ii) the number (density of states) of spin-up channels within the energy interval close to the Fermi level, $E_\text{F}\pm0.12$\,eV, Fig.\,\ref{fig:cell}(a1). Similar results, regarding the spin-polarized currents, are expected for other (X/\can/X)$_n$ widths ($n$) with X\,=\,H, and F. We believe that these findings are important for the development and design of spintronic devices in 2D materials, in particular, the ones based on the hydrogenation or fluorination \can\, electrenes.

\section{Summary and Conclusions}

In summary, we have performed a first-principles investigation of  the electronic properties of the hydrogenated and fluorinated \can\, electrenes,   X/\can/X, with X\,=\,H and  F.  A net magnetic moment arises as a result of the partial occupation of the nitrogen atoms between the passivated Ca atoms, followed by the suppression of the NFE states on the \can\ surface upon such passivation. In the sequence, we have considered lateral heterostructures composed of a combination of pristine and passivated \can\, MLs, (\can)(X/\can/X). Those heterostructures are characterized by the emergence of 1D channels of spin-polarized  NFE states on the  (\can) surface confined by the passivated (X/\can/X) stripes. Moreover, we found that the passivated X/\can/X regions act as a spin-dependent scattering center, giving rise to spin-polarized electronic current. These findings show that the passivation of \can\, electrenes is an appealing route to designing spin-polarized electronic devices based on NFEs on  2D platforms.

\begin{acknowledgments}

The authors acknowledge financial support from the Brazilian agencies CNPq, 
CAPES, and FAPEMIG, INCT-Nanomateriais de Carbono, Rede-Mineira de Materiais 2D, and the CENAPAD-SP and Laborat{\'o}rio Nacional de 
Computa{\c{c}}{\~a}o Cient{\'i}fica (LNCC-SCAFMat2) for computer time.

\end{acknowledgments}
 \section{Appendix}

\begin{figure}[h]
    \includegraphics[width=8.5cm]{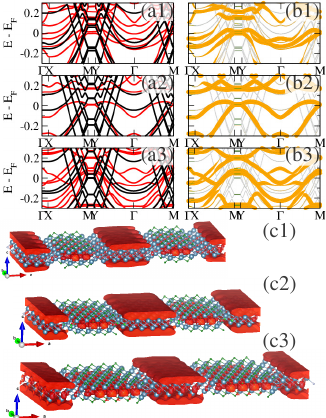}
    \caption{The electronic band structure and orbital projected electronic band structure on the fluorinated region of ((\can)$_6$/(F/\can/F)$_3$ (a1,b1), ((\can)$_6$/(F/\can/F)$_4$ (a2,b2), and ((\can)$_6$/(F/\can/F)$_5$ (a3,b3); the spin-density distribution of ((\can)$_6$/(F/\can/F)$_3$ (c1), ((\can)$_6$/(F/\can/F)$_4$ (c2), and ((\can)$_6$/(F/\can/F)$_5$ (c3)}
    \label{apen2}
\end{figure}

 In Figure \ref{apen2} we show the electronic band structure and the respective orbital projected electronic band structure of the (\can)$_6$/(F/\can/F)$_3$ (a1)-(b1),  (\can)$_6$/(F/\can/F)$_4$ (a2)-(b2), and (\can)$_6$/(F/\can/F)$_5$ (a3)-(b3). In Figure \ref{apen2}(c1)-(c3), we present the spin-density distribution of (\can)$_6$/(F/\can/F)$_n$.
 
\bibliography{trilha.bib}

\begin{thebibliography}{44}%
\makeatletter
\providecommand \@ifxundefined [1]{%
 \@ifx{#1\undefined}
}%
\providecommand \@ifnum [1]{%
 \ifnum #1\expandafter \@firstoftwo
 \else \expandafter \@secondoftwo
 \fi
}%
\providecommand \@ifx [1]{%
 \ifx #1\expandafter \@firstoftwo
 \else \expandafter \@secondoftwo
 \fi
}%
\providecommand \natexlab [1]{#1}%
\providecommand \enquote  [1]{``#1''}%
\providecommand \bibnamefont  [1]{#1}%
\providecommand \bibfnamefont [1]{#1}%
\providecommand \citenamefont [1]{#1}%
\providecommand \href@noop [0]{\@secondoftwo}%
\providecommand \href [0]{\begingroup \@sanitize@url \@href}%
\providecommand \@href[1]{\@@startlink{#1}\@@href}%
\providecommand \@@href[1]{\endgroup#1\@@endlink}%
\providecommand \@sanitize@url [0]{\catcode `\\12\catcode `\$12\catcode
  `\&12\catcode `\#12\catcode `\^12\catcode `\_12\catcode `\%12\relax}%
\providecommand \@@startlink[1]{}%
\providecommand \@@endlink[0]{}%
\providecommand \url  [0]{\begingroup\@sanitize@url \@url }%
\providecommand \@url [1]{\endgroup\@href {#1}{\urlprefix }}%
\providecommand \urlprefix  [0]{URL }%
\providecommand \Eprint [0]{\href }%
\providecommand \doibase [0]{http://dx.doi.org/}%
\providecommand \selectlanguage [0]{\@gobble}%
\providecommand \bibinfo  [0]{\@secondoftwo}%
\providecommand \bibfield  [0]{\@secondoftwo}%
\providecommand \translation [1]{[#1]}%
\providecommand \BibitemOpen [0]{}%
\providecommand \bibitemStop [0]{}%
\providecommand \bibitemNoStop [0]{.\EOS\space}%
\providecommand \EOS [0]{\spacefactor3000\relax}%
\providecommand \BibitemShut  [1]{\csname bibitem#1\endcsname}%
\let\auto@bib@innerbib\@empty
\bibitem [{\citenamefont {Oh}\ \emph {et~al.}(2016)\citenamefont {Oh},
  \citenamefont {Kang}, \citenamefont {Kim}, \citenamefont {Sinn},
  \citenamefont {Han}, \citenamefont {Chang}, \citenamefont {Park},
  \citenamefont {Kim}, \citenamefont {Min}, \citenamefont {Kim} \emph
  {et~al.}}]{oh2016evidence}%
  \BibitemOpen
  \bibfield  {author} {\bibinfo {author} {\bibfnamefont {J.~S.}\ \bibnamefont
  {Oh}}, \bibinfo {author} {\bibfnamefont {C.-J.}\ \bibnamefont {Kang}},
  \bibinfo {author} {\bibfnamefont {Y.~J.}\ \bibnamefont {Kim}}, \bibinfo
  {author} {\bibfnamefont {S.}~\bibnamefont {Sinn}}, \bibinfo {author}
  {\bibfnamefont {M.}~\bibnamefont {Han}}, \bibinfo {author} {\bibfnamefont
  {Y.~J.}\ \bibnamefont {Chang}}, \bibinfo {author} {\bibfnamefont {B.-G.}\
  \bibnamefont {Park}}, \bibinfo {author} {\bibfnamefont {S.~W.}\ \bibnamefont
  {Kim}}, \bibinfo {author} {\bibfnamefont {B.~I.}\ \bibnamefont {Min}},
  \bibinfo {author} {\bibfnamefont {H.-D.}\ \bibnamefont {Kim}},  \emph
  {et~al.},\ }\href@noop {} {\bibfield  {journal} {\bibinfo  {journal} {Journal
  of the American Chemical Society}\ }\textbf {\bibinfo {volume} {138}},\
  \bibinfo {pages} {2496} (\bibinfo {year} {2016})}\BibitemShut {NoStop}%
\bibitem [{\citenamefont {Sui}\ \emph {et~al.}(2019)\citenamefont {Sui},
  \citenamefont {Wang},\ and\ \citenamefont {Duan}}]{sui2019prediction}%
  \BibitemOpen
  \bibfield  {author} {\bibinfo {author} {\bibfnamefont {X.}~\bibnamefont
  {Sui}}, \bibinfo {author} {\bibfnamefont {J.}~\bibnamefont {Wang}}, \ and\
  \bibinfo {author} {\bibfnamefont {W.}~\bibnamefont {Duan}},\ }\href@noop {}
  {\bibfield  {journal} {\bibinfo  {journal} {The Journal of Physical Chemistry
  C}\ }\textbf {\bibinfo {volume} {123}},\ \bibinfo {pages} {5003} (\bibinfo
  {year} {2019})}\BibitemShut {NoStop}%
\bibitem [{\citenamefont {Lee}\ \emph {et~al.}(2013)\citenamefont {Lee},
  \citenamefont {Kim}, \citenamefont {Toda}, \citenamefont {Matsuishi},\ and\
  \citenamefont {Hosono}}]{lee2013dicalcium}%
  \BibitemOpen
  \bibfield  {author} {\bibinfo {author} {\bibfnamefont {K.}~\bibnamefont
  {Lee}}, \bibinfo {author} {\bibfnamefont {S.~W.}\ \bibnamefont {Kim}},
  \bibinfo {author} {\bibfnamefont {Y.}~\bibnamefont {Toda}}, \bibinfo {author}
  {\bibfnamefont {S.}~\bibnamefont {Matsuishi}}, \ and\ \bibinfo {author}
  {\bibfnamefont {H.}~\bibnamefont {Hosono}},\ }\href@noop {} {\bibfield
  {journal} {\bibinfo  {journal} {Nature}\ }\textbf {\bibinfo {volume} {494}},\
  \bibinfo {pages} {336} (\bibinfo {year} {2013})}\BibitemShut {NoStop}%
\bibitem [{\citenamefont {Druffel}\ \emph {et~al.}(2017)\citenamefont
  {Druffel}, \citenamefont {Woomer}, \citenamefont {Kuntz}, \citenamefont
  {Pawlik},\ and\ \citenamefont {Warren}}]{druffel2017electrons}%
  \BibitemOpen
  \bibfield  {author} {\bibinfo {author} {\bibfnamefont {D.~L.}\ \bibnamefont
  {Druffel}}, \bibinfo {author} {\bibfnamefont {A.~H.}\ \bibnamefont {Woomer}},
  \bibinfo {author} {\bibfnamefont {K.~L.}\ \bibnamefont {Kuntz}}, \bibinfo
  {author} {\bibfnamefont {J.~T.}\ \bibnamefont {Pawlik}}, \ and\ \bibinfo
  {author} {\bibfnamefont {S.~C.}\ \bibnamefont {Warren}},\ }\href@noop {}
  {\bibfield  {journal} {\bibinfo  {journal} {Journal of Materials Chemistry
  C}\ }\textbf {\bibinfo {volume} {5}},\ \bibinfo {pages} {11196} (\bibinfo
  {year} {2017})}\BibitemShut {NoStop}%
\bibitem [{\citenamefont {Wang}\ \emph {et~al.}(2018)\citenamefont {Wang},
  \citenamefont {Li}, \citenamefont {Shen}, \citenamefont {Guo}, \citenamefont
  {Li}, \citenamefont {Zhao}, \citenamefont {Fang},\ and\ \citenamefont
  {Yang}}]{wang2018ultralow}%
  \BibitemOpen
  \bibfield  {author} {\bibinfo {author} {\bibfnamefont {J.}~\bibnamefont
  {Wang}}, \bibinfo {author} {\bibfnamefont {L.}~\bibnamefont {Li}}, \bibinfo
  {author} {\bibfnamefont {Z.}~\bibnamefont {Shen}}, \bibinfo {author}
  {\bibfnamefont {P.}~\bibnamefont {Guo}}, \bibinfo {author} {\bibfnamefont
  {M.}~\bibnamefont {Li}}, \bibinfo {author} {\bibfnamefont {B.}~\bibnamefont
  {Zhao}}, \bibinfo {author} {\bibfnamefont {L.}~\bibnamefont {Fang}}, \ and\
  \bibinfo {author} {\bibfnamefont {L.}~\bibnamefont {Yang}},\ }\href@noop {}
  {\bibfield  {journal} {\bibinfo  {journal} {Materials}\ }\textbf {\bibinfo
  {volume} {11}},\ \bibinfo {pages} {2462} (\bibinfo {year}
  {2018})}\BibitemShut {NoStop}%
\bibitem [{\citenamefont {Druffel}\ \emph {et~al.}(2016)\citenamefont
  {Druffel}, \citenamefont {Kuntz}, \citenamefont {Woomer}, \citenamefont
  {Alcorn}, \citenamefont {Hu}, \citenamefont {Donley},\ and\ \citenamefont
  {Warren}}]{druffel2016experimental}%
  \BibitemOpen
  \bibfield  {author} {\bibinfo {author} {\bibfnamefont {D.~L.}\ \bibnamefont
  {Druffel}}, \bibinfo {author} {\bibfnamefont {K.~L.}\ \bibnamefont {Kuntz}},
  \bibinfo {author} {\bibfnamefont {A.~H.}\ \bibnamefont {Woomer}}, \bibinfo
  {author} {\bibfnamefont {F.~M.}\ \bibnamefont {Alcorn}}, \bibinfo {author}
  {\bibfnamefont {J.}~\bibnamefont {Hu}}, \bibinfo {author} {\bibfnamefont
  {C.~L.}\ \bibnamefont {Donley}}, \ and\ \bibinfo {author} {\bibfnamefont
  {S.~C.}\ \bibnamefont {Warren}},\ }\href@noop {} {\bibfield  {journal}
  {\bibinfo  {journal} {Journal of the American Chemical Society}\ }\textbf
  {\bibinfo {volume} {138}},\ \bibinfo {pages} {16089} (\bibinfo {year}
  {2016})}\BibitemShut {NoStop}%
\bibitem [{\citenamefont {Dale}\ and\ \citenamefont
  {Johnson}(2017)}]{dale2017ionic}%
  \BibitemOpen
  \bibfield  {author} {\bibinfo {author} {\bibfnamefont {S.~G.}\ \bibnamefont
  {Dale}}\ and\ \bibinfo {author} {\bibfnamefont {E.~R.}\ \bibnamefont
  {Johnson}},\ }\href@noop {} {\bibfield  {journal} {\bibinfo  {journal}
  {Physical Chemistry Chemical Physics}\ }\textbf {\bibinfo {volume} {19}},\
  \bibinfo {pages} {27343} (\bibinfo {year} {2017})}\BibitemShut {NoStop}%
\bibitem [{\citenamefont {Hirayama}\ \emph {et~al.}(2018)\citenamefont
  {Hirayama}, \citenamefont {Matsuishi}, \citenamefont {Hosono},\ and\
  \citenamefont {Murakami}}]{hirayama2018electrides}%
  \BibitemOpen
  \bibfield  {author} {\bibinfo {author} {\bibfnamefont {M.}~\bibnamefont
  {Hirayama}}, \bibinfo {author} {\bibfnamefont {S.}~\bibnamefont {Matsuishi}},
  \bibinfo {author} {\bibfnamefont {H.}~\bibnamefont {Hosono}}, \ and\ \bibinfo
  {author} {\bibfnamefont {S.}~\bibnamefont {Murakami}},\ }\href@noop {}
  {\bibfield  {journal} {\bibinfo  {journal} {Physical Review X}\ }\textbf
  {\bibinfo {volume} {8}},\ \bibinfo {pages} {031067} (\bibinfo {year}
  {2018})}\BibitemShut {NoStop}%
\bibitem [{\citenamefont {Huang}\ \emph {et~al.}(2018)\citenamefont {Huang},
  \citenamefont {Jin}, \citenamefont {Zhang},\ and\ \citenamefont
  {Liu}}]{huang2018topological}%
  \BibitemOpen
  \bibfield  {author} {\bibinfo {author} {\bibfnamefont {H.}~\bibnamefont
  {Huang}}, \bibinfo {author} {\bibfnamefont {K.-H.}\ \bibnamefont {Jin}},
  \bibinfo {author} {\bibfnamefont {S.}~\bibnamefont {Zhang}}, \ and\ \bibinfo
  {author} {\bibfnamefont {F.}~\bibnamefont {Liu}},\ }\href@noop {} {\bibfield
  {journal} {\bibinfo  {journal} {Nano letters}\ }\textbf {\bibinfo {volume}
  {18}},\ \bibinfo {pages} {1972} (\bibinfo {year} {2018})}\BibitemShut
  {NoStop}%
\bibitem [{\citenamefont {Park}\ \emph {et~al.}(2018)\citenamefont {Park},
  \citenamefont {Kim},\ and\ \citenamefont {Yoon}}]{park2018first}%
  \BibitemOpen
  \bibfield  {author} {\bibinfo {author} {\bibfnamefont {C.}~\bibnamefont
  {Park}}, \bibinfo {author} {\bibfnamefont {S.~W.}\ \bibnamefont {Kim}}, \
  and\ \bibinfo {author} {\bibfnamefont {M.}~\bibnamefont {Yoon}},\ }\href@noop
  {} {\bibfield  {journal} {\bibinfo  {journal} {Physical review letters}\
  }\textbf {\bibinfo {volume} {120}},\ \bibinfo {pages} {026401} (\bibinfo
  {year} {2018})}\BibitemShut {NoStop}%
\bibitem [{\citenamefont {Zhu}\ \emph {et~al.}(2019)\citenamefont {Zhu},
  \citenamefont {Wang}, \citenamefont {Qu}, \citenamefont {Wang}, \citenamefont
  {Frolov}, \citenamefont {Chen},\ and\ \citenamefont
  {Zhu}}]{zhu2019computational}%
  \BibitemOpen
  \bibfield  {author} {\bibinfo {author} {\bibfnamefont {S.-C.}\ \bibnamefont
  {Zhu}}, \bibinfo {author} {\bibfnamefont {L.}~\bibnamefont {Wang}}, \bibinfo
  {author} {\bibfnamefont {J.-Y.}\ \bibnamefont {Qu}}, \bibinfo {author}
  {\bibfnamefont {J.-J.}\ \bibnamefont {Wang}}, \bibinfo {author}
  {\bibfnamefont {T.}~\bibnamefont {Frolov}}, \bibinfo {author} {\bibfnamefont
  {X.-Q.}\ \bibnamefont {Chen}}, \ and\ \bibinfo {author} {\bibfnamefont
  {Q.}~\bibnamefont {Zhu}},\ }\href@noop {} {\bibfield  {journal} {\bibinfo
  {journal} {Physical Review Materials}\ }\textbf {\bibinfo {volume} {3}},\
  \bibinfo {pages} {024205} (\bibinfo {year} {2019})}\BibitemShut {NoStop}%
\bibitem [{\citenamefont {Zhang}\ \emph {et~al.}(2018)\citenamefont {Zhang},
  \citenamefont {Guo}, \citenamefont {Jin}, \citenamefont {Dai},\ and\
  \citenamefont {Liu}}]{zhang2018intermetallic}%
  \BibitemOpen
  \bibfield  {author} {\bibinfo {author} {\bibfnamefont {X.}~\bibnamefont
  {Zhang}}, \bibinfo {author} {\bibfnamefont {R.}~\bibnamefont {Guo}}, \bibinfo
  {author} {\bibfnamefont {L.}~\bibnamefont {Jin}}, \bibinfo {author}
  {\bibfnamefont {X.}~\bibnamefont {Dai}}, \ and\ \bibinfo {author}
  {\bibfnamefont {G.}~\bibnamefont {Liu}},\ }\href@noop {} {\bibfield
  {journal} {\bibinfo  {journal} {Journal of Materials Chemistry C}\ }\textbf
  {\bibinfo {volume} {6}},\ \bibinfo {pages} {575} (\bibinfo {year}
  {2018})}\BibitemShut {NoStop}%
\bibitem [{\citenamefont {Zhang}\ \emph {et~al.}(2019)\citenamefont {Zhang},
  \citenamefont {Fu}, \citenamefont {Jin}, \citenamefont {Dai}, \citenamefont
  {Liu},\ and\ \citenamefont {Yao}}]{zhang2019topological}%
  \BibitemOpen
  \bibfield  {author} {\bibinfo {author} {\bibfnamefont {X.}~\bibnamefont
  {Zhang}}, \bibinfo {author} {\bibfnamefont {B.}~\bibnamefont {Fu}}, \bibinfo
  {author} {\bibfnamefont {L.}~\bibnamefont {Jin}}, \bibinfo {author}
  {\bibfnamefont {X.}~\bibnamefont {Dai}}, \bibinfo {author} {\bibfnamefont
  {G.}~\bibnamefont {Liu}}, \ and\ \bibinfo {author} {\bibfnamefont
  {Y.}~\bibnamefont {Yao}},\ }\href@noop {} {\bibfield  {journal} {\bibinfo
  {journal} {The Journal of Physical Chemistry C}\ }\textbf {\bibinfo {volume}
  {123}},\ \bibinfo {pages} {25871} (\bibinfo {year} {2019})}\BibitemShut
  {NoStop}%
\bibitem [{\citenamefont {Liu}\ \emph {et~al.}(2019)\citenamefont {Liu},
  \citenamefont {Wang}, \citenamefont {Yi}, \citenamefont {Kim}, \citenamefont
  {Park},\ and\ \citenamefont {Cho}}]{liu2019theoretical}%
  \BibitemOpen
  \bibfield  {author} {\bibinfo {author} {\bibfnamefont {L.}~\bibnamefont
  {Liu}}, \bibinfo {author} {\bibfnamefont {C.}~\bibnamefont {Wang}}, \bibinfo
  {author} {\bibfnamefont {S.}~\bibnamefont {Yi}}, \bibinfo {author}
  {\bibfnamefont {D.~K.}\ \bibnamefont {Kim}}, \bibinfo {author} {\bibfnamefont
  {C.~H.}\ \bibnamefont {Park}}, \ and\ \bibinfo {author} {\bibfnamefont
  {J.-H.}\ \bibnamefont {Cho}},\ }\href@noop {} {\bibfield  {journal} {\bibinfo
   {journal} {Physical Review B}\ }\textbf {\bibinfo {volume} {99}},\ \bibinfo
  {pages} {220401} (\bibinfo {year} {2019})}\BibitemShut {NoStop}%
\bibitem [{\citenamefont {Hou}\ \emph {et~al.}(2016)\citenamefont {Hou},
  \citenamefont {Tu},\ and\ \citenamefont {Chen}}]{hou2016two}%
  \BibitemOpen
  \bibfield  {author} {\bibinfo {author} {\bibfnamefont {J.}~\bibnamefont
  {Hou}}, \bibinfo {author} {\bibfnamefont {K.}~\bibnamefont {Tu}}, \ and\
  \bibinfo {author} {\bibfnamefont {Z.}~\bibnamefont {Chen}},\ }\href@noop {}
  {\bibfield  {journal} {\bibinfo  {journal} {The Journal of Physical Chemistry
  C}\ }\textbf {\bibinfo {volume} {120}},\ \bibinfo {pages} {18473} (\bibinfo
  {year} {2016})}\BibitemShut {NoStop}%
\bibitem [{\citenamefont {Chen}\ \emph {et~al.}(2017)\citenamefont {Chen},
  \citenamefont {Bai}, \citenamefont {Li}, \citenamefont {Li}, \citenamefont
  {Wang}, \citenamefont {Ni}, \citenamefont {Liu}, \citenamefont {Wu},
  \citenamefont {Yao},\ and\ \citenamefont {Wu}}]{chen2017multilayered}%
  \BibitemOpen
  \bibfield  {author} {\bibinfo {author} {\bibfnamefont {G.}~\bibnamefont
  {Chen}}, \bibinfo {author} {\bibfnamefont {Y.}~\bibnamefont {Bai}}, \bibinfo
  {author} {\bibfnamefont {H.}~\bibnamefont {Li}}, \bibinfo {author}
  {\bibfnamefont {Y.}~\bibnamefont {Li}}, \bibinfo {author} {\bibfnamefont
  {Z.}~\bibnamefont {Wang}}, \bibinfo {author} {\bibfnamefont {Q.}~\bibnamefont
  {Ni}}, \bibinfo {author} {\bibfnamefont {L.}~\bibnamefont {Liu}}, \bibinfo
  {author} {\bibfnamefont {F.}~\bibnamefont {Wu}}, \bibinfo {author}
  {\bibfnamefont {Y.}~\bibnamefont {Yao}}, \ and\ \bibinfo {author}
  {\bibfnamefont {C.}~\bibnamefont {Wu}},\ }\href@noop {} {\bibfield  {journal}
  {\bibinfo  {journal} {ACS applied materials \& interfaces}\ }\textbf
  {\bibinfo {volume} {9}},\ \bibinfo {pages} {6666} (\bibinfo {year}
  {2017})}\BibitemShut {NoStop}%
\bibitem [{\citenamefont {Kocabas}\ \emph {et~al.}(2018)\citenamefont
  {Kocabas}, \citenamefont {Ozden}, \citenamefont {Demiroglu}, \citenamefont
  {{\c{C}}ak{\i}r},\ and\ \citenamefont {Sevik}}]{2018determination}%
  \BibitemOpen
  \bibfield  {author} {\bibinfo {author} {\bibfnamefont {T.}~\bibnamefont
  {Kocabas}}, \bibinfo {author} {\bibfnamefont {A.}~\bibnamefont {Ozden}},
  \bibinfo {author} {\bibfnamefont {I.}~\bibnamefont {Demiroglu}}, \bibinfo
  {author} {\bibfnamefont {D.}~\bibnamefont {{\c{C}}ak{\i}r}}, \ and\ \bibinfo
  {author} {\bibfnamefont {C.}~\bibnamefont {Sevik}},\ }\href@noop {}
  {\bibfield  {journal} {\bibinfo  {journal} {The journal of physical chemistry
  letters}\ }\textbf {\bibinfo {volume} {9}},\ \bibinfo {pages} {4267}
  (\bibinfo {year} {2018})}\BibitemShut {NoStop}%
\bibitem [{\citenamefont {Tang}\ \emph {et~al.}(2018)\citenamefont {Tang},
  \citenamefont {Wan}, \citenamefont {Gao}, \citenamefont {Muraba},
  \citenamefont {Qin}, \citenamefont {Yan}, \citenamefont {Chen}, \citenamefont
  {Hu}, \citenamefont {Zhang}, \citenamefont {Wu} \emph
  {et~al.}}]{tang2018metal}%
  \BibitemOpen
  \bibfield  {author} {\bibinfo {author} {\bibfnamefont {H.}~\bibnamefont
  {Tang}}, \bibinfo {author} {\bibfnamefont {B.}~\bibnamefont {Wan}}, \bibinfo
  {author} {\bibfnamefont {B.}~\bibnamefont {Gao}}, \bibinfo {author}
  {\bibfnamefont {Y.}~\bibnamefont {Muraba}}, \bibinfo {author} {\bibfnamefont
  {Q.}~\bibnamefont {Qin}}, \bibinfo {author} {\bibfnamefont {B.}~\bibnamefont
  {Yan}}, \bibinfo {author} {\bibfnamefont {P.}~\bibnamefont {Chen}}, \bibinfo
  {author} {\bibfnamefont {Q.}~\bibnamefont {Hu}}, \bibinfo {author}
  {\bibfnamefont {D.}~\bibnamefont {Zhang}}, \bibinfo {author} {\bibfnamefont
  {L.}~\bibnamefont {Wu}},  \emph {et~al.},\ }\href@noop {} {\bibfield
  {journal} {\bibinfo  {journal} {Advanced Science}\ }\textbf {\bibinfo
  {volume} {5}},\ \bibinfo {pages} {1800666} (\bibinfo {year}
  {2018})}\BibitemShut {NoStop}%
\bibitem [{\citenamefont {Inoshita}\ \emph {et~al.}(2017)\citenamefont
  {Inoshita}, \citenamefont {Tsukada}, \citenamefont {Saito},\ and\
  \citenamefont {Hosono}}]{inoshita2017probing}%
  \BibitemOpen
  \bibfield  {author} {\bibinfo {author} {\bibfnamefont {T.}~\bibnamefont
  {Inoshita}}, \bibinfo {author} {\bibfnamefont {M.}~\bibnamefont {Tsukada}},
  \bibinfo {author} {\bibfnamefont {S.}~\bibnamefont {Saito}}, \ and\ \bibinfo
  {author} {\bibfnamefont {H.}~\bibnamefont {Hosono}},\ }\href@noop {}
  {\bibfield  {journal} {\bibinfo  {journal} {Physical Review B}\ }\textbf
  {\bibinfo {volume} {96}},\ \bibinfo {pages} {245303} (\bibinfo {year}
  {2017})}\BibitemShut {NoStop}%
\bibitem [{\citenamefont {Choi}\ \emph {et~al.}(2021)\citenamefont {Choi},
  \citenamefont {Kim}, \citenamefont {Choi}, \citenamefont {Cha},\ and\
  \citenamefont {Hong}}]{choi2021electronic}%
  \BibitemOpen
  \bibfield  {author} {\bibinfo {author} {\bibfnamefont {C.-G.}\ \bibnamefont
  {Choi}}, \bibinfo {author} {\bibfnamefont {J.}~\bibnamefont {Kim}}, \bibinfo
  {author} {\bibfnamefont {H.-K.}\ \bibnamefont {Choi}}, \bibinfo {author}
  {\bibfnamefont {J.}~\bibnamefont {Cha}}, \ and\ \bibinfo {author}
  {\bibfnamefont {S.}~\bibnamefont {Hong}},\ }\href@noop {} {\bibfield
  {journal} {\bibinfo  {journal} {Current Applied Physics}\ }\textbf {\bibinfo
  {volume} {28}},\ \bibinfo {pages} {13} (\bibinfo {year} {2021})}\BibitemShut
  {NoStop}%
\bibitem [{\citenamefont {Wu}\ and\ \citenamefont
  {Yao}(2020)}]{wu2020intriguing}%
  \BibitemOpen
  \bibfield  {author} {\bibinfo {author} {\bibfnamefont {C.-W.}\ \bibnamefont
  {Wu}}\ and\ \bibinfo {author} {\bibfnamefont {D.-X.}\ \bibnamefont {Yao}},\
  }\href@noop {} {\bibfield  {journal} {\bibinfo  {journal} {Journal of
  Magnetism and Magnetic Materials}\ }\textbf {\bibinfo {volume} {493}},\
  \bibinfo {pages} {165727} (\bibinfo {year} {2020})}\BibitemShut {NoStop}%
\bibitem [{\citenamefont {Souza}\ \emph {et~al.}(2020)\citenamefont {Souza},
  \citenamefont {Padilha},\ and\ \citenamefont {Miwa}}]{souza2020structural}%
  \BibitemOpen
  \bibfield  {author} {\bibinfo {author} {\bibfnamefont {P.~H.}\ \bibnamefont
  {Souza}}, \bibinfo {author} {\bibfnamefont {J.~E.}\ \bibnamefont {Padilha}},
  \ and\ \bibinfo {author} {\bibfnamefont {R.~H.}\ \bibnamefont {Miwa}},\
  }\href@noop {} {\bibfield  {journal} {\bibinfo  {journal} {The Journal of
  Physical Chemistry C}\ }\textbf {\bibinfo {volume} {124}},\ \bibinfo {pages}
  {14706} (\bibinfo {year} {2020})}\BibitemShut {NoStop}%
\bibitem [{\citenamefont {Qiu}\ \emph {et~al.}(2019)\citenamefont {Qiu},
  \citenamefont {Zhang}, \citenamefont {Lu},\ and\ \citenamefont
  {Liu}}]{hidrox}%
  \BibitemOpen
  \bibfield  {author} {\bibinfo {author} {\bibfnamefont {X.-L.}\ \bibnamefont
  {Qiu}}, \bibinfo {author} {\bibfnamefont {J.-F.}\ \bibnamefont {Zhang}},
  \bibinfo {author} {\bibfnamefont {Z.-Y.}\ \bibnamefont {Lu}}, \ and\ \bibinfo
  {author} {\bibfnamefont {K.}~\bibnamefont {Liu}},\ }\href@noop {} {\bibfield
  {journal} {\bibinfo  {journal} {The Journal of Physical Chemistry C}\
  }\textbf {\bibinfo {volume} {123}},\ \bibinfo {pages} {24698} (\bibinfo
  {year} {2019})}\BibitemShut {NoStop}%
\bibitem [{\citenamefont {Wang}\ \emph {et~al.}(2019)\citenamefont {Wang},
  \citenamefont {Li}, \citenamefont {Zhang}, \citenamefont {Sun}, \citenamefont
  {Han}, \citenamefont {Niu}, \citenamefont {Zhong}, \citenamefont {Qu},\ and\
  \citenamefont {Yang}}]{wang2019first}%
  \BibitemOpen
  \bibfield  {author} {\bibinfo {author} {\bibfnamefont {D.}~\bibnamefont
  {Wang}}, \bibinfo {author} {\bibfnamefont {H.}~\bibnamefont {Li}}, \bibinfo
  {author} {\bibfnamefont {L.}~\bibnamefont {Zhang}}, \bibinfo {author}
  {\bibfnamefont {Z.}~\bibnamefont {Sun}}, \bibinfo {author} {\bibfnamefont
  {D.}~\bibnamefont {Han}}, \bibinfo {author} {\bibfnamefont {L.}~\bibnamefont
  {Niu}}, \bibinfo {author} {\bibfnamefont {X.}~\bibnamefont {Zhong}}, \bibinfo
  {author} {\bibfnamefont {X.}~\bibnamefont {Qu}}, \ and\ \bibinfo {author}
  {\bibfnamefont {L.}~\bibnamefont {Yang}},\ }\href@noop {} {\bibfield
  {journal} {\bibinfo  {journal} {Applied Surface Science}\ }\textbf {\bibinfo
  {volume} {478}},\ \bibinfo {pages} {459} (\bibinfo {year}
  {2019})}\BibitemShut {NoStop}%
\bibitem [{\citenamefont {Gong}\ \emph {et~al.}(2014)\citenamefont {Gong},
  \citenamefont {Lin}, \citenamefont {Wang}, \citenamefont {Shi}, \citenamefont
  {Lei}, \citenamefont {Lin}, \citenamefont {Zou}, \citenamefont {Ye},
  \citenamefont {Vajtai}, \citenamefont {Yakobson} \emph
  {et~al.}}]{gong2014vertical}%
  \BibitemOpen
  \bibfield  {author} {\bibinfo {author} {\bibfnamefont {Y.}~\bibnamefont
  {Gong}}, \bibinfo {author} {\bibfnamefont {J.}~\bibnamefont {Lin}}, \bibinfo
  {author} {\bibfnamefont {X.}~\bibnamefont {Wang}}, \bibinfo {author}
  {\bibfnamefont {G.}~\bibnamefont {Shi}}, \bibinfo {author} {\bibfnamefont
  {S.}~\bibnamefont {Lei}}, \bibinfo {author} {\bibfnamefont {Z.}~\bibnamefont
  {Lin}}, \bibinfo {author} {\bibfnamefont {X.}~\bibnamefont {Zou}}, \bibinfo
  {author} {\bibfnamefont {G.}~\bibnamefont {Ye}}, \bibinfo {author}
  {\bibfnamefont {R.}~\bibnamefont {Vajtai}}, \bibinfo {author} {\bibfnamefont
  {B.~I.}\ \bibnamefont {Yakobson}},  \emph {et~al.},\ }\href@noop {}
  {\bibfield  {journal} {\bibinfo  {journal} {Nature materials}\ }\textbf
  {\bibinfo {volume} {13}},\ \bibinfo {pages} {1135} (\bibinfo {year}
  {2014})}\BibitemShut {NoStop}%
\bibitem [{\citenamefont {Fang}\ \emph {et~al.}(2018)\citenamefont {Fang},
  \citenamefont {Tian}, \citenamefont {Sheng}, \citenamefont {Yang},
  \citenamefont {Lu}, \citenamefont {Wang}, \citenamefont {Zhang},
  \citenamefont {Zhang}, \citenamefont {Yan},\ and\ \citenamefont
  {Hua}}]{fang2018chemical}%
  \BibitemOpen
  \bibfield  {author} {\bibinfo {author} {\bibfnamefont {X.}~\bibnamefont
  {Fang}}, \bibinfo {author} {\bibfnamefont {Q.}~\bibnamefont {Tian}}, \bibinfo
  {author} {\bibfnamefont {Y.}~\bibnamefont {Sheng}}, \bibinfo {author}
  {\bibfnamefont {G.}~\bibnamefont {Yang}}, \bibinfo {author} {\bibfnamefont
  {N.}~\bibnamefont {Lu}}, \bibinfo {author} {\bibfnamefont {J.}~\bibnamefont
  {Wang}}, \bibinfo {author} {\bibfnamefont {X.}~\bibnamefont {Zhang}},
  \bibinfo {author} {\bibfnamefont {Y.}~\bibnamefont {Zhang}}, \bibinfo
  {author} {\bibfnamefont {X.}~\bibnamefont {Yan}}, \ and\ \bibinfo {author}
  {\bibfnamefont {B.}~\bibnamefont {Hua}},\ }\href@noop {} {\bibfield
  {journal} {\bibinfo  {journal} {Superlattices and Microstructures}\ }\textbf
  {\bibinfo {volume} {123}},\ \bibinfo {pages} {323} (\bibinfo {year}
  {2018})}\BibitemShut {NoStop}%
\bibitem [{\citenamefont {Ci}\ \emph {et~al.}(2010)\citenamefont {Ci},
  \citenamefont {Song}, \citenamefont {Jin}, \citenamefont {Jariwala},
  \citenamefont {Wu}, \citenamefont {Li}, \citenamefont {Srivastava},
  \citenamefont {Wang}, \citenamefont {Storr}, \citenamefont {Balicas} \emph
  {et~al.}}]{ci2010atomic}%
  \BibitemOpen
  \bibfield  {author} {\bibinfo {author} {\bibfnamefont {L.}~\bibnamefont
  {Ci}}, \bibinfo {author} {\bibfnamefont {L.}~\bibnamefont {Song}}, \bibinfo
  {author} {\bibfnamefont {C.}~\bibnamefont {Jin}}, \bibinfo {author}
  {\bibfnamefont {D.}~\bibnamefont {Jariwala}}, \bibinfo {author}
  {\bibfnamefont {D.}~\bibnamefont {Wu}}, \bibinfo {author} {\bibfnamefont
  {Y.}~\bibnamefont {Li}}, \bibinfo {author} {\bibfnamefont {A.}~\bibnamefont
  {Srivastava}}, \bibinfo {author} {\bibfnamefont {Z.}~\bibnamefont {Wang}},
  \bibinfo {author} {\bibfnamefont {K.}~\bibnamefont {Storr}}, \bibinfo
  {author} {\bibfnamefont {L.}~\bibnamefont {Balicas}},  \emph {et~al.},\
  }\href@noop {} {\bibfield  {journal} {\bibinfo  {journal} {Nature materials}\
  }\textbf {\bibinfo {volume} {9}},\ \bibinfo {pages} {430} (\bibinfo {year}
  {2010})}\BibitemShut {NoStop}%
\bibitem [{\citenamefont {Liu}\ \emph {et~al.}(2018)\citenamefont {Liu},
  \citenamefont {Zhang}, \citenamefont {Wang}, \citenamefont {Yakobson},\ and\
  \citenamefont {Hersam}}]{liu2018intermixing}%
  \BibitemOpen
  \bibfield  {author} {\bibinfo {author} {\bibfnamefont {X.}~\bibnamefont
  {Liu}}, \bibinfo {author} {\bibfnamefont {Z.}~\bibnamefont {Zhang}}, \bibinfo
  {author} {\bibfnamefont {L.}~\bibnamefont {Wang}}, \bibinfo {author}
  {\bibfnamefont {B.~I.}\ \bibnamefont {Yakobson}}, \ and\ \bibinfo {author}
  {\bibfnamefont {M.~C.}\ \bibnamefont {Hersam}},\ }\href@noop {} {\bibfield
  {journal} {\bibinfo  {journal} {Nature materials}\ }\textbf {\bibinfo
  {volume} {17}},\ \bibinfo {pages} {783} (\bibinfo {year} {2018})}\BibitemShut
  {NoStop}%
\bibitem [{\citenamefont {Silvestre}\ \emph {et~al.}(2019)\citenamefont
  {Silvestre}, \citenamefont {Scopel},\ and\ \citenamefont
  {Miwa}}]{silvestre2019electronic}%
  \BibitemOpen
  \bibfield  {author} {\bibinfo {author} {\bibfnamefont {G.}~\bibnamefont
  {Silvestre}}, \bibinfo {author} {\bibfnamefont {W.~L.}\ \bibnamefont
  {Scopel}}, \ and\ \bibinfo {author} {\bibfnamefont {R.}~\bibnamefont
  {Miwa}},\ }\href@noop {} {\bibfield  {journal} {\bibinfo  {journal}
  {Nanoscale}\ }\textbf {\bibinfo {volume} {11}},\ \bibinfo {pages} {17894}
  (\bibinfo {year} {2019})}\BibitemShut {NoStop}%
\bibitem [{\citenamefont {Khabibrakhmanov}\ and\ \citenamefont
  {Sorokin}(2022)}]{khabibrakhmanov2022electronic}%
  \BibitemOpen
  \bibfield  {author} {\bibinfo {author} {\bibfnamefont {A.~I.}\ \bibnamefont
  {Khabibrakhmanov}}\ and\ \bibinfo {author} {\bibfnamefont {P.~B.}\
  \bibnamefont {Sorokin}},\ }\href@noop {} {\bibfield  {journal} {\bibinfo
  {journal} {Nanoscale}\ }\textbf {\bibinfo {volume} {14}},\ \bibinfo {pages}
  {4131} (\bibinfo {year} {2022})}\BibitemShut {NoStop}%
\bibitem [{\citenamefont {Varlamova}\ \emph {et~al.}(2022)\citenamefont
  {Varlamova}, \citenamefont {Erohin}, \citenamefont {Larionov},\ and\
  \citenamefont {Sorokin}}]{varlamova2022diamane}%
  \BibitemOpen
  \bibfield  {author} {\bibinfo {author} {\bibfnamefont {L.~A.}\ \bibnamefont
  {Varlamova}}, \bibinfo {author} {\bibfnamefont {S.~V.}\ \bibnamefont
  {Erohin}}, \bibinfo {author} {\bibfnamefont {K.~V.}\ \bibnamefont
  {Larionov}}, \ and\ \bibinfo {author} {\bibfnamefont {P.~B.}\ \bibnamefont
  {Sorokin}},\ }\href@noop {} {\bibfield  {journal} {\bibinfo  {journal} {The
  Journal of Physical Chemistry Letters}\ }\textbf {\bibinfo {volume} {13}},\
  \bibinfo {pages} {11383} (\bibinfo {year} {2022})}\BibitemShut {NoStop}%
\bibitem [{\citenamefont {Hohenberg}\ and\ \citenamefont {Kohn}(1964)}]{hk}%
  \BibitemOpen
  \bibfield  {author} {\bibinfo {author} {\bibfnamefont {P.}~\bibnamefont
  {Hohenberg}}\ and\ \bibinfo {author} {\bibfnamefont {W.}~\bibnamefont
  {Kohn}},\ }\href@noop {} {\bibfield  {journal} {\bibinfo  {journal} {Physical
  review}\ }\textbf {\bibinfo {volume} {136}},\ \bibinfo {pages} {B864}
  (\bibinfo {year} {1964})}\BibitemShut {NoStop}%
\bibitem [{\citenamefont {Kresse}\ and\ \citenamefont
  {Furthm{\"u}ller}(1996{\natexlab{a}})}]{vasp1}%
  \BibitemOpen
  \bibfield  {author} {\bibinfo {author} {\bibfnamefont {G.}~\bibnamefont
  {Kresse}}\ and\ \bibinfo {author} {\bibfnamefont {J.}~\bibnamefont
  {Furthm{\"u}ller}},\ }\href@noop {} {\bibfield  {journal} {\bibinfo
  {journal} {Computational materials science}\ }\textbf {\bibinfo {volume}
  {6}},\ \bibinfo {pages} {15} (\bibinfo {year}
  {1996}{\natexlab{a}})}\BibitemShut {NoStop}%
\bibitem [{\citenamefont {Kresse}\ and\ \citenamefont
  {Furthm{\"u}ller}(1996{\natexlab{b}})}]{vasp2}%
  \BibitemOpen
  \bibfield  {author} {\bibinfo {author} {\bibfnamefont {G.}~\bibnamefont
  {Kresse}}\ and\ \bibinfo {author} {\bibfnamefont {J.}~\bibnamefont
  {Furthm{\"u}ller}},\ }\href@noop {} {\bibfield  {journal} {\bibinfo
  {journal} {Physical review B}\ }\textbf {\bibinfo {volume} {54}},\ \bibinfo
  {pages} {11169} (\bibinfo {year} {1996}{\natexlab{b}})}\BibitemShut {NoStop}%
\bibitem [{\citenamefont {Perdew}\ \emph {et~al.}(1996)\citenamefont {Perdew},
  \citenamefont {Burke},\ and\ \citenamefont {Ernzerhof}}]{pbe}%
  \BibitemOpen
  \bibfield  {author} {\bibinfo {author} {\bibfnamefont {J.~P.}\ \bibnamefont
  {Perdew}}, \bibinfo {author} {\bibfnamefont {K.}~\bibnamefont {Burke}}, \
  and\ \bibinfo {author} {\bibfnamefont {M.}~\bibnamefont {Ernzerhof}},\
  }\href@noop {} {\bibfield  {journal} {\bibinfo  {journal} {Physical review
  letters}\ }\textbf {\bibinfo {volume} {77}},\ \bibinfo {pages} {3865}
  (\bibinfo {year} {1996})}\BibitemShut {NoStop}%
\bibitem [{\citenamefont {Monkhorst}\ and\ \citenamefont
  {Pack}(1976)}]{kpoints}%
  \BibitemOpen
  \bibfield  {author} {\bibinfo {author} {\bibfnamefont {H.~J.}\ \bibnamefont
  {Monkhorst}}\ and\ \bibinfo {author} {\bibfnamefont {J.~D.}\ \bibnamefont
  {Pack}},\ }\href@noop {} {\bibfield  {journal} {\bibinfo  {journal} {Physical
  review B}\ }\textbf {\bibinfo {volume} {13}},\ \bibinfo {pages} {5188}
  (\bibinfo {year} {1976})}\BibitemShut {NoStop}%
\bibitem [{\citenamefont {Soler}\ \emph {et~al.}(2002)\citenamefont {Soler},
  \citenamefont {Artacho}, \citenamefont {Gale}, \citenamefont {Garc{\'\i}a},
  \citenamefont {Junquera}, \citenamefont {Ordej{\'o}n},\ and\ \citenamefont
  {S{\'a}nchez-Portal}}]{siesta}%
  \BibitemOpen
  \bibfield  {author} {\bibinfo {author} {\bibfnamefont {J.~M.}\ \bibnamefont
  {Soler}}, \bibinfo {author} {\bibfnamefont {E.}~\bibnamefont {Artacho}},
  \bibinfo {author} {\bibfnamefont {J.~D.}\ \bibnamefont {Gale}}, \bibinfo
  {author} {\bibfnamefont {A.}~\bibnamefont {Garc{\'\i}a}}, \bibinfo {author}
  {\bibfnamefont {J.}~\bibnamefont {Junquera}}, \bibinfo {author}
  {\bibfnamefont {P.}~\bibnamefont {Ordej{\'o}n}}, \ and\ \bibinfo {author}
  {\bibfnamefont {D.}~\bibnamefont {S{\'a}nchez-Portal}},\ }\href@noop {}
  {\bibfield  {journal} {\bibinfo  {journal} {Journal of Physics: Condensed
  Matter}\ }\textbf {\bibinfo {volume} {14}},\ \bibinfo {pages} {2745}
  (\bibinfo {year} {2002})}\BibitemShut {NoStop}%
\bibitem [{\citenamefont {Brandbyge}\ \emph {et~al.}(2002)\citenamefont
  {Brandbyge}, \citenamefont {Mozos}, \citenamefont {Ordej{\'o}n},
  \citenamefont {Taylor},\ and\ \citenamefont {Stokbro}}]{transiesta}%
  \BibitemOpen
  \bibfield  {author} {\bibinfo {author} {\bibfnamefont {M.}~\bibnamefont
  {Brandbyge}}, \bibinfo {author} {\bibfnamefont {J.-L.}\ \bibnamefont
  {Mozos}}, \bibinfo {author} {\bibfnamefont {P.}~\bibnamefont {Ordej{\'o}n}},
  \bibinfo {author} {\bibfnamefont {J.}~\bibnamefont {Taylor}}, \ and\ \bibinfo
  {author} {\bibfnamefont {K.}~\bibnamefont {Stokbro}},\ }\href@noop {}
  {\bibfield  {journal} {\bibinfo  {journal} {Physical Review B}\ }\textbf
  {\bibinfo {volume} {65}},\ \bibinfo {pages} {165401} (\bibinfo {year}
  {2002})}\BibitemShut {NoStop}%
\bibitem [{\citenamefont {Artacho}(1999)}]{artacho1999sanchez}%
  \BibitemOpen
  \bibfield  {author} {\bibinfo {author} {\bibfnamefont {D.}~\bibnamefont
  {Artacho}},\ }\href@noop {} {\bibfield  {journal} {\bibinfo  {journal} {Phys.
  Status Solidi B}\ }\textbf {\bibinfo {volume} {215}},\ \bibinfo {pages} {809}
  (\bibinfo {year} {1999})}\BibitemShut {NoStop}%
\bibitem [{\citenamefont {Leenaerts}\ \emph {et~al.}(2010)\citenamefont
  {Leenaerts}, \citenamefont {Peelaers}, \citenamefont {Hern{\'a}ndez-Nieves},
  \citenamefont {Partoens},\ and\ \citenamefont
  {Peeters}}]{leenaerts2010first}%
  \BibitemOpen
  \bibfield  {author} {\bibinfo {author} {\bibfnamefont {O.}~\bibnamefont
  {Leenaerts}}, \bibinfo {author} {\bibfnamefont {H.}~\bibnamefont {Peelaers}},
  \bibinfo {author} {\bibfnamefont {A.}~\bibnamefont {Hern{\'a}ndez-Nieves}},
  \bibinfo {author} {\bibfnamefont {B.}~\bibnamefont {Partoens}}, \ and\
  \bibinfo {author} {\bibfnamefont {F.}~\bibnamefont {Peeters}},\ }\href@noop
  {} {\bibfield  {journal} {\bibinfo  {journal} {Physical Review B}\ }\textbf
  {\bibinfo {volume} {82}},\ \bibinfo {pages} {195436} (\bibinfo {year}
  {2010})}\BibitemShut {NoStop}%
\bibitem [{\citenamefont {Cort{\'e}s-del R{\'\i}o}\ \emph
  {et~al.}(2020)\citenamefont {Cort{\'e}s-del R{\'\i}o}, \citenamefont
  {Mallet}, \citenamefont {Gonz{\'a}lez-Herrero}, \citenamefont {Lado},
  \citenamefont {Fern{\'a}ndez-Rossier}, \citenamefont
  {G{\'o}mez-Rodr{\'\i}guez}, \citenamefont {Veuillen},\ and\ \citenamefont
  {Brihuega}}]{cortes2020quantum}%
  \BibitemOpen
  \bibfield  {author} {\bibinfo {author} {\bibfnamefont {E.}~\bibnamefont
  {Cort{\'e}s-del R{\'\i}o}}, \bibinfo {author} {\bibfnamefont
  {P.}~\bibnamefont {Mallet}}, \bibinfo {author} {\bibfnamefont
  {H.}~\bibnamefont {Gonz{\'a}lez-Herrero}}, \bibinfo {author} {\bibfnamefont
  {J.~L.}\ \bibnamefont {Lado}}, \bibinfo {author} {\bibfnamefont
  {J.}~\bibnamefont {Fern{\'a}ndez-Rossier}}, \bibinfo {author} {\bibfnamefont
  {J.~M.}\ \bibnamefont {G{\'o}mez-Rodr{\'\i}guez}}, \bibinfo {author}
  {\bibfnamefont {J.-Y.}\ \bibnamefont {Veuillen}}, \ and\ \bibinfo {author}
  {\bibfnamefont {I.}~\bibnamefont {Brihuega}},\ }\href@noop {} {\bibfield
  {journal} {\bibinfo  {journal} {Advanced Materials}\ }\textbf {\bibinfo
  {volume} {32}},\ \bibinfo {pages} {2001119} (\bibinfo {year}
  {2020})}\BibitemShut {NoStop}%
\bibitem [{\citenamefont {Chernozatonskii}\ \emph {et~al.}(2007)\citenamefont
  {Chernozatonskii}, \citenamefont {Sorokin},\ and\ \citenamefont
  {Br{\"u}ning}}]{chernozatonskii2007two}%
  \BibitemOpen
  \bibfield  {author} {\bibinfo {author} {\bibfnamefont {L.~A.}\ \bibnamefont
  {Chernozatonskii}}, \bibinfo {author} {\bibfnamefont {P.~B.}\ \bibnamefont
  {Sorokin}}, \ and\ \bibinfo {author} {\bibfnamefont {J.~W.}\ \bibnamefont
  {Br{\"u}ning}},\ }\href@noop {} {\bibfield  {journal} {\bibinfo  {journal}
  {Applied Physics Letters}\ }\textbf {\bibinfo {volume} {91}},\ \bibinfo
  {pages} {183103} (\bibinfo {year} {2007})}\BibitemShut {NoStop}%
\bibitem [{\citenamefont {Wu}\ \emph {et~al.}(2009)\citenamefont {Wu},
  \citenamefont {Wu}, \citenamefont {Gao},\ and\ \citenamefont
  {Zeng}}]{wu2009materials}%
  \BibitemOpen
  \bibfield  {author} {\bibinfo {author} {\bibfnamefont {M.}~\bibnamefont
  {Wu}}, \bibinfo {author} {\bibfnamefont {X.}~\bibnamefont {Wu}}, \bibinfo
  {author} {\bibfnamefont {Y.}~\bibnamefont {Gao}}, \ and\ \bibinfo {author}
  {\bibfnamefont {X.~C.}\ \bibnamefont {Zeng}},\ }\href@noop {} {\bibfield
  {journal} {\bibinfo  {journal} {Applied Physics Letters}\ }\textbf {\bibinfo
  {volume} {94}},\ \bibinfo {pages} {223111} (\bibinfo {year}
  {2009})}\BibitemShut {NoStop}%
\bibitem [{\citenamefont {Singh}\ and\ \citenamefont
  {Yakobson}(2009)}]{singh2009electronics}%
  \BibitemOpen
  \bibfield  {author} {\bibinfo {author} {\bibfnamefont {A.~K.}\ \bibnamefont
  {Singh}}\ and\ \bibinfo {author} {\bibfnamefont {B.~I.}\ \bibnamefont
  {Yakobson}},\ }\href@noop {} {\bibfield  {journal} {\bibinfo  {journal} {Nano
  Letters}\ }\textbf {\bibinfo {volume} {9}},\ \bibinfo {pages} {1540}
  (\bibinfo {year} {2009})}\BibitemShut {NoStop}%
\end{thebibliography}%

\end{document}